# Cosmological functional renormalisation group, extended Galilean invariance and the sweeping effect

Alaric Erschfeld[*] and Stefan Floerchinger[†]
*Institut für Theoretische Physik, Ruprecht-Karls-Universität Heidelberg*
*Philosophenweg 16, D-69120 Heidelberg, Germany*

The functional renormalisation group is employed to study the non-linear regime of late-time cosmic structure formation. This framework naturally allows for non-perturbative approximation schemes, usually guided by underlying symmetries or a truncation of the theory space. An extended symmetry that is related to Galilean invariance is studied and corresponding Ward identities are derived. These are used to obtain (formally) closed renormalisation group flow equations for two-point correlation functions in the limit of large wave numbers (small scales). The flow equations are analytically solved in an approximation that is connected to the 'sweeping effect' previously described in the context of fluid turbulence.

## I. INTRODUCTION

One of the primary goals of contemporary cosmology is to describe how dark matter evolves under the influence of gravity in order to understand the observed large-scale structure of the Universe. As observations probe increasingly smaller scales, there is a genuine need for methods that allow to describe gravitational dynamics at smaller scales.

Cosmic large-scale structure formation can be described in terms of kinetic theory, where dark matter is modelled by self-gravitating classical point particles on an expanding space-time. On large scales (of the order of megaparsecs), deviations from a homogeneous and isotropic background are relatively small and fluctuations are well described by linear theory. In order to take non-linearities into account, the dynamics of fluctuations can be solved perturbatively around their linear solutions. Formally, this is sensible as long as the variance of the dark matter mass density fluctuation field is small. Since deviations from the homogeneous and isotropic background get large at cosmically late times and small scales, perturbation theory is no longer applicable.

Cosmological perturbation theory is usually studied from the Lagrangian or Eulerian point of view [1] and in what follows, the latter is adopted. In order to investigate non-linear structure formation beyond standard Eulerian perturbation theory, different methods have been proposed. These include various resummation schemes [2–9], two-particle irreducible methods [10–12], direct interaction approximations [13–15], the renormalisation group [16–19], effective theories [20–22], higher-order perturbation theory and extensions thereof [23–25] and kinetic field theory [26–28], to name only a few. In the following, the functional renormalisation group is studied, which has proven to be very useful for various non-perturbative phenomena in quantum field theories and statistical mechanics [29, 30].

The functional renormalisation group employed here describes conceptually how a theory without initial state fluctuations changes when the latter are gradually included. This provides a framework in which correlation functions can be computed in a generically non-perturbative way. A caveat is that in most cases the functional renormalisation group equations can not be solved exactly, since an infinite hierarchy of coupled functional differential equations is involved. In order to obtain a solvable system, one therefore often resorts to approximation schemes such as a derivative or vertex expansion, effectively truncating the theory space. Another route is to use the underlying symmetries of the field theory to restrict the space of possible solutions. The symmetries relate to generalised Ward identities that can sometimes be used to solve the renormalisation group in certain limits or sectors of the theory.

The paper is organised as follows. In section II the kinetic theory description of dark matter is reviewed and the statistical field theory due to stochastic initial states is motivated. In section III an action-based functional formalism and the functional renormalisation group are introduced. In section IV the symmetries of the theory, in particular mass conservation and extended Galilean invariance, are used to derive related Ward identities and restrict the form of the effective action. In section V the Ward identities related to extended Galilean invariance are used in the limit of large wave numbers to (formally) close the two-point correlation function flow equations. These are analytically solved in the infrared regime of the flow equations and the relation to the sweeping effect is investigated. Finally, some conclusions are drawn in section VI.

## II. KINETIC THEORY DESCRIPTION OF DARK MATTER

In the Newtonian limit of a kinetic theory description, dark matter is often modelled by an ensem-ble of self-gravitating, collisionless point particles of mass $m$ that evolve on an expanding (flat) Friedmann–Lemaître–Robertson–Walker space-time. The state of the theory is described by a one-particle phase-space distribution function $f(\tau, \boldsymbol{x}, \boldsymbol{p})$ whose dynamics is governed by the Vlasov–Poisson equations [31],

$$\begin{aligned} \partial_\tau f + \frac{p_i}{am}\,\partial_i f - am\,\partial_i\phi\,\frac{\partial f}{\partial p_i} &= 0\ , \\ \partial_i\partial_i\phi = \tfrac{3}{2}\mathcal{H}^2\varOmega_\mathrm{m}&\left[\int_{\mathbb{R}^3}\frac{\mathrm{d}^3p}{(2\pi)^3}\,f - 1\right]. \end{aligned} \tag{1}$$

[*] erschfeld@thphys.uni-heidelberg.de
[†] floerchinger@thphys.uni-heidelberg.de

Here, $\tau$ is conformal time, $\boldsymbol{x}$ are comoving coordinates and $\boldsymbol{p} = am\,\mathrm{d}\boldsymbol{x}/\mathrm{d}\tau$ are the corresponding conjugate momenta.[1] The former two are related to cosmic time and proper physical coordinates by $\mathrm{d}t = a(\tau)\,\mathrm{d}\tau$ and $\boldsymbol{r} = a\boldsymbol{x}$, respectively. Here $a$ is the scale factor parametrising the expansion of space, $\mathcal{H} = \dot{a}/a$ is the conformal Hubble function and $\varOmega_\mathrm{m}$ is the (time-dependent) dark matter density parameter. Finally, the distribution function is normalised to

$$\int_{\mathbb{R}^3}\frac{\mathrm{d}^3p}{(2\pi)^3}\,\langle f(\tau, \boldsymbol{x}, \boldsymbol{p})\rangle = 1\ , \tag{2}$$

and $\phi(\tau, \boldsymbol{x})$ is the (peculiar) Newtonian gravitational potential.[2]

Since the system of equations (1) is non-local and non-linear, it is quite difficult to solve for the full distribution function. However, one is often rather interested in moments or cumulants with respect to the momentum argument, the complete set of which fully characterise the distribution function. Although being two sides of the same coin, it is often preferential to work in terms of cumulants since they are the connected part of the moments. The first few cumulants are the (logarithmic) density contrast $\ln(1 + \delta(\tau, \boldsymbol{x}))$, quantifying the local mass density deviation relative to the mean, the velocity $u_i(\tau, \boldsymbol{x})$ and the velocity dispersion tensor $\sigma_{ij}(\tau, \boldsymbol{x})$.

The Vlasov–Poisson equations can be cast into an infinite tower of coupled evolution equations for the cumulants [32]. Qualitatively, the evolution equation of the $n$th-order cumulant is given by [33]

$$\begin{aligned} \partial_\tau c^{(n)} + n\mathcal{H}c^{(n)} + \partial_{\boldsymbol{x}}c^{(n+1)} + \sum_{l=0}^{n}\binom{n}{l}\,c^{(l+1)}\partial_{\boldsymbol{x}}c^{(n-l)} \\ + \delta_{n1}\partial_{\boldsymbol{x}}\phi = 0\ , \end{aligned} \tag{3}$$

where the tensorial structure of the cumulants is suppressed and $\delta_{ij}$ denotes the Kronecker delta. Together with Poisson's equation,

$$\partial_i\partial_i\phi = \tfrac{3}{2}\mathcal{H}^2\varOmega_\mathrm{m}\delta\ , \tag{4}$$

they form a closed system of equations.

From a practical point of view it is rather useless to keep the full cumulant expansion since one has to deal with an infinite amount of equations, being equivalent to solving the full Vlasov–Poisson equations (1). In order to obtain a solvable system of equations, one therefore often turns to approximations, in particular to truncations of the cumulant expansion.

The simplest non-trivial approximation is obtained by truncating the cumulant expansion at second order. The momentum dependence of the distribution function in the so-called *single-stream approximation* is degenerate,

$$f = (1 + \delta)\,(2\pi)^3\delta(\boldsymbol{p} - am\boldsymbol{u})\ , \tag{5}$$

such that momentum $p_i$ is directly related to the velocity field $u_i$ at each instant in time and point in space. Here and in the following, $\delta(\boldsymbol{p} - am\boldsymbol{u})$ denotes the (three-dimensional) Dirac delta function. The single-stream approximation models dark matter as a perfect pressureless fluid described in terms of the density and velocity field only. The relevant evolution equations are the continuity and Euler's equations, which are obtained at zeroth and first order from equation (3), respectively.

While the single-stream approximation successfully describes early-time and very large-scale structure formation, it fails to accurately capture gravitational dynamics at later times and smaller scales where physical processes related to higher-order cumulants become important. More specifically, during gravitational collapse the trajectories of dark matter particles cross in position space, a phenomenon known as *shell-crossing*. After shell-crossing the velocity field is multi-valued so that the multiple roots of the distribution function (5) generate a non-trivial velocity dispersion tensor. In turn all higher-order cumulants are sourced which indicates the breakdown of the perfect pressureless fluid model [34].

A simple extension beyond the single-stream approximation is the inclusion of the velocity dispersion tensor and a truncation of the cumulant expansion thereafter [35]. Correspondingly, momentum is normal distributed and the velocity dispersion tensor is the covariance matrix,

$$\begin{aligned} f = \frac{(1+\delta)\,(2\pi)^{\frac{3}{2}}}{a^3m^3\det(\sigma)^{\frac{1}{2}}}\exp\Big\{-\frac{1}{2}\Big(\frac{p_i}{am} - u_i\Big)(\sigma^{-1})_{ij} \\ \times\Big(\frac{p_j}{am} - u_j\Big)\Big\}\ . \end{aligned} \tag{6}$$

[1] Spatial vector components are referred to by indices from the middle of the Latin alphabet, while boldface symbols denote the corresponding vector. Partial derivatives with respect to conformal time are often abbreviated as an overdot while those with respect to comoving coordinate components are referred to by $\partial_i$. Einstein's summation convention is employed, where repeated indices in a single term are summed over, although vectors and covectors are not distinguished, as is common in flat space. Finally, function arguments are often suppressed for the sake of brevity.

[2] Expectation values $\langle\ldots\rangle$ are taken either as ensembles averages over cosmic histories with stochastic initial conditions or as sample averages over large spatial volumes in a single cosmic history. This is discussed in more detail at the end of this section.

The velocity dispersion tensor regularises the momentum delta function of the single-stream approximation (5), which is recovered in the limit $\sigma_{ij} \to 0$.[3] In addition to the continuity and Cauchy momentum equations one has to include an evolution equation for the velocity dispersion tensor, which is obtained from equation (3) at second order. Although the distribution function (6) can not capture shell-crossing microscopically, it supports the average motion of a multi-stream flow.

While the single-stream approximation is mathematically self-consistent, at least in the absence of shell-crossing, higher-order cumulants are naturally generated by non-linear terms for non-vanishing velocity dispersion [34]. This makes a rigorous justification for a truncation of the cumulant expansion rather difficult. From a physical point of view one could argue that the distribution function (6) naturally supports a Maxwell–Boltzmann distribution of momenta such as expected for non-relativistic particles that decouple thermally in the primordial Universe [36] or virialised clumps of dark matter, at least in simple halo models [37]. On the other hand, it seems obvious that the distribution function (6) has a natural range of scales where it is applicable but ultimately breaks down at sufficiently small scales, as is likely for any description including only a finite amount of cumulants.

More generally, one can include cumulants up to some desired order $n$ and truncate the expansion thereafter. In the case of $n > 2$ the distribution function can no longer be explicitly reconstructed and the generation of higher-order cumulants is similar to the case $n = 2$ so that a self-consistent truncation is not possible.

In the following, it is convenient to introduce the compact notation

$$\psi_a = (\delta,\ u_i,\ \sigma_{ij},\ \dots)\ , \tag{7}$$

where the desired field content is included into the multiplet $\psi_a(\tau, \boldsymbol{x})$. The index $a$ carries any additional substructure of the fields, like representations of the rotation group, such as for the velocity or velocity dispersion field, and is summed over for repeated indices. The equations of motion (3) can be cast into the form

$$\partial_\tau \psi_a(\tau,\boldsymbol{x}) + \int_{\boldsymbol{x}'} \Omega_{ab}(\tau,\boldsymbol{x}-\boldsymbol{x}')\,\psi_b(\tau,\boldsymbol{x}') + \int_{\boldsymbol{x}',\boldsymbol{x}''} \gamma_{abc}(\boldsymbol{x}-\boldsymbol{x}',\boldsymbol{x}-\boldsymbol{x}'')\,\psi_b(\tau,\boldsymbol{x}')\,\psi_c(\tau,\boldsymbol{x}'') = 0\ , \tag{8}$$

where

$$\Omega(\tau,\boldsymbol{x}-\boldsymbol{x}') = \begin{array}{c} \\ \delta \\ u_i \\ \sigma_{ij} \\ \\ \end{array} \begin{array}{c} \begin{array}{cccc} \delta & u_j & \sigma_{kl} & \end{array} \\ \begin{pmatrix} 0 & \partial_j\delta(\boldsymbol{x}-\boldsymbol{x}') & 0 & \cdots \\ O_i(\tau,\boldsymbol{x}-\boldsymbol{x}') & \mathcal{H}\delta_{ij} & \delta_{ik}\partial_l\delta(\boldsymbol{x}-\boldsymbol{x}') & \cdots \\ 0 & 0 & 2\mathcal{H}\delta_{ik}\delta_{jl} & \cdots \\ \vdots & \vdots & \vdots & \ddots \end{pmatrix} \end{array}\ , \tag{9}$$

and the fields above and to the left of the matrix are displayed as an orientation. The matrix $\Omega_{ab}$ is local and of upper triangular form except for the velocity-density component that is determined by the solution of Poisson's equation (4) and given by the operator

$$O_i(\tau,\boldsymbol{x}-\boldsymbol{x}') = \frac{3}{2}\,\mathcal{H}^2\Omega_\mathrm{m}\,\frac{1}{4\pi}\,\frac{x_i - x_i'}{|\boldsymbol{x}-\boldsymbol{x}'|^3}\ . \tag{10}$$

The diagonal entries of $\Omega_{ab}$ are the Hubble drag terms of the corresponding cumulants, while the entries above the diagonal are derivative couplings to the next higher cumulant. The vertices follow from the non-linear terms in equation (3), act as gradients and are defined such that they exhibit the symmetry $\gamma_{abc}(\boldsymbol{x}-\boldsymbol{x}',\boldsymbol{x}-\boldsymbol{x}'') = \gamma_{acb}(\boldsymbol{x}-\boldsymbol{x}'',\boldsymbol{x}-\boldsymbol{x}')$.[4]

In cosmology one is often less interested in the exact solution of the Vlasov–Poisson equations (1) but rather in the statistical properties of the system for random initial conditions. Most inflation models predict the initial state of the theory to be very near to a Gaussian random field subject to the *cosmological principle*, i.e. the primordial cosmic fields are statistically homogeneous and isotropic in space. The employed statistical field theory can be understood as describing an ensemble of cosmic histories with stochastic initial conditions or, equivalently for this purpose, a sample of large spatial volumes of a single cosmic history.[5]

---

[3] Since the velocity dispersion tensor is positive definite by construction, the distribution function (6) exists, except for trivial degenerate cases.

[4] Since the last term of the sum in equation (3) is $c^{(n+1)}\partial_{\boldsymbol{x}}\ln(1+\delta)$, which is non-polynomial in the fields, equation (8) can not capture the exact cumulant dynamics. In the following, these terms are approximated as $c^{(n+1)}\partial_{\boldsymbol{x}}\delta$, which can be understood as a vertex expansion around the (vanishing) density contrast mean field.

[5] The *fair sample hypothesis* is most often assumed in cosmology

In a statistic field theory description of dark matter one is typically interested in expectation values of cosmic field products, formally similar to moments and cumulants of the phase-space distribution function. Assuming that the symmetries of the cosmological principle are realised statistically, one is concerned with expectation values such as the mean field,

$$\langle \psi_a(\tau, \boldsymbol{x}) \rangle = \Psi_a(\tau) \ , \tag{11}$$

the covariance function,

$$\langle \psi_a(\tau, \boldsymbol{x})\, \psi_b(\tau', \boldsymbol{x}') \rangle_{\mathrm{c}} = C_{ab}(\tau, \tau', \boldsymbol{x} - \boldsymbol{x}') \ , \tag{12}$$

and other higher-order correlation functions.[6] While the single-stream approximation features no mean fields, an example where it becomes relevant is for the truncation (6) where the trace of the velocity dispersion tensor mean field couples non-trivially with fluctuations [39, 40].

In the following, it is convenient to study correlations also in Fourier space.[7] One of the central objects studied in cosmology is the power spectral density of the covariance function,

$$C_{ab}(\tau, \tau', \boldsymbol{x} - \boldsymbol{x}') = \int_{\boldsymbol{q}} \mathrm{e}^{\mathrm{i}\boldsymbol{q}\cdot(\boldsymbol{x}-\boldsymbol{x}')}\, P_{ab}(\tau, \tau', \boldsymbol{q}) \ , \tag{13}$$

which is invariant under parity transformations, $P_{ab}(\tau, \tau', -\boldsymbol{q}) = P_{ab}(\tau, \tau', \boldsymbol{q})$, due to the statistical isotropy symmetry. Since the initial distribution of cosmic fields is Gaussian, it is fully characterised in terms of the mean fields and power spectra. The subsequent nonlinear evolution in time naturally drives the distribution away from its Gaussian shape so that higher-order correlations are generated. These can also be studied in terms of the bi- and trispectrum, which quantify the skewness and kurtosis of the distribution.

## III. COSMOLOGICAL FUNCTIONAL RENORMALISATION GROUP

In this section the Martin–Siggia–Rose/Janssen–de Dominicis formalism [41–44] as well as the functional renormalisation group for stochastic dynamics is reviewed. Although already established in cosmology [17, 19], some aspects are recapitulated in order to introduce notation and the inclusion of non-vanishing mean fields is treated, such as needed for a description beyond the single-stream approximation.

The initial Gaussian state of the fields (7) is completely characterised by the mean fields $\Psi_a^{\mathrm{in}}$ and the covariance functions $C_{ab}^{\mathrm{in}}(\boldsymbol{x} - \boldsymbol{x}')$, and the subsequent evolution of correlations is determined by the equations of motion (8). Following the Martin–Siggia–Rose formalism, one introduces a conjugate set of so-called response fields $\hat{\psi}_a(\tau, \boldsymbol{x})$ which are utilised in the Janssen–de Dominicis formalism to constructs a functional integral representation with the generating functional

$$Z[J, \hat{J}; \Psi^{\mathrm{in}}, C^{\mathrm{in}}] = \int \mathcal{D}\psi \int \mathcal{D}\hat{\psi}\ \mathrm{e}^{-S + J_{\boldsymbol{A}}\psi_{\boldsymbol{A}}} \ . \tag{14}$$

Here, $S$ is the bare action

$$\begin{aligned} S[\psi, \hat{\psi}; \Psi^{\mathrm{in}}, C^{\mathrm{in}}] = &-\mathrm{i} \int_{\tau, \boldsymbol{x}, \boldsymbol{x}'} \hat{\psi}_a(\tau, \boldsymbol{x}) \Big[\partial_\tau \delta_{ab}\, \delta(\boldsymbol{x} - \boldsymbol{x}') + \Omega_{ab}(\tau, \boldsymbol{x} - \boldsymbol{x}')\Big] \psi_b(\tau, \boldsymbol{x}') \\ &- \mathrm{i} \int_{\tau, \boldsymbol{x}, \boldsymbol{x}', \boldsymbol{x}''} \hat{\psi}_a(\tau, \boldsymbol{x})\, \gamma_{abc}(\boldsymbol{x} - \boldsymbol{x}', \boldsymbol{x} - \boldsymbol{x}'')\, \psi_b(\tau, \boldsymbol{x}')\, \psi_c(\tau, \boldsymbol{x}'') \\ &+ \int_{\boldsymbol{x}, \boldsymbol{x}'} \hat{\psi}_a(\tau_{\mathrm{in}}, \boldsymbol{x}) \Big[\mathrm{i}\, \delta(\boldsymbol{x} - \boldsymbol{x}')\, \Psi_a^{\mathrm{in}} + \tfrac{1}{2}\, C_{ab}^{\mathrm{in}}(\boldsymbol{x} - \boldsymbol{x}')\, \hat{\psi}_b(\tau_{\mathrm{in}}, \boldsymbol{x}')\Big] \ , \end{aligned} \tag{15}$$

and the integral measures $\mathcal{D}\psi$ and $\mathcal{D}\hat{\psi}$ are understood as the continuum limit of integrals over field values on a lattice in time and space.[8]

and closely related to the concept of ergodicity where ensemble averages are equal to sample averages over an infinite volume [38]. Rigorously, ergodicity holds for statistically homogeneous Gaussian random fields with continuous power spectral density but is rather difficult to proof for more general cases.

[6] Here, $\langle \ldots \rangle_{\mathrm{c}}$ denotes the connected part of the expectation value.

[7] The Fourier transform convention

$$f(\boldsymbol{q}) = \int_{\boldsymbol{x}} \mathrm{e}^{-\mathrm{i}\boldsymbol{q}\cdot\boldsymbol{x}}\, f(\boldsymbol{x}) \ , \qquad f(\boldsymbol{x}) = \int_{\boldsymbol{q}} \mathrm{e}^{\mathrm{i}\boldsymbol{q}\cdot\boldsymbol{x}}\, f(\boldsymbol{q}) \ ,$$

is employed, where $\boldsymbol{q} \cdot \boldsymbol{x} = q_i x_i$ denotes the standard Euclidean inner product and the modulus is $q = |\boldsymbol{q}|$. Integrals over the entire domain are abbreviated as

$$\int_{\boldsymbol{x}} = \int_{\mathbb{R}^3} \mathrm{d}^3 x \ , \qquad \int_{\boldsymbol{q}} = \int_{\mathbb{R}^3} \frac{\mathrm{d}^3 q}{(2\pi)^3} \ ,$$

and wave vector delta functions are denoted as $\delta\!\!\!{}^-(\boldsymbol{q}) = (2\pi)^3\, \delta(\boldsymbol{q})$.

[8] The construction assumes a unique solution of the equations of

Capital letters from the beginning of the Latin alphabet denote DeWitt indices, e.g. $A = (a, \tau, \boldsymbol{x})$, which are summed and integrated over for discrete and continuous variables respectively, while boldface indices additionally comprise the physical-response field structure, e.g. $\psi_{\boldsymbol{A}} = (\psi_A, \widehat{\psi}_A)$.

Correlation functions are obtained from the generating functional (14) by applying functional derivatives with respect to the source currents $J_{\boldsymbol{A}}$,

$$\frac{Z^{(n)}_{\boldsymbol{A}_1 \dots \boldsymbol{A}_n}}{Z} = \frac{1}{Z} \frac{\delta^n Z}{\delta J_{\boldsymbol{A}_1} \dots \delta J_{\boldsymbol{A}_n}} = \langle \psi_{\boldsymbol{A}_1} \dots \psi_{\boldsymbol{A}_n} \rangle \,, \qquad (16)$$

where physical correlations are obtained at vanishing source currents and are said to be 'on the equations of motion'.[9]

Since the action (15) describes an interacting theory, it is naturally difficult to compute correlation functions in a non-perturbative manner. In standard cosmological perturbation theory one computes correlation functions around the non-interacting theory in orders of the linear power spectrum. Unfortunately, the regime where perturbation theory is applicable is restricted to rather early times or large scales since the variance of the linear power spectrum grows large in the opposite of either of these two regimes. Going deeper into the ultraviolet at late times therefore requires non-perturbative methods, such as the functional renormalisation group.

In order to use the functional renormalisation group to compute correlation functions one regularises the generating functional (14) by adding a term that is bilinear in the fields to the bare action,

$$S_k = S + \tfrac{1}{2}\, \psi_{\boldsymbol{A}} R_{k,\boldsymbol{AB}}\, \psi_{\boldsymbol{B}} \,. \qquad (17)$$

The regulator $R_k$ depends on the renormalisation group scale $k$ and suppresses fluctuations within the functional integral, either in the infrared or ultraviolet (or both), such that one obtains a regulated generating functional $Z_k$.

For the purpose of this work it is particularly convenient to use $R_k$ to modify the initial power spectrum only and different values of $k$ correspond to physical situations with different initial power spectra. One can organise this such that one limit ($k \to 0$ for the ultraviolet regulator (30) employed here) corresponds to a vanishing initial power spectrum, implying that the theory is free of fluctuations. One can then study how correlations change when the renormalisation group scale $k$ is altered so that more fluctuations are taken into account.

To this end, the generating functional of connected correlation functions $W_k = \ln(Z_k)$ is introduced. Mean fields are obtained from first order functional derivatives, which on the equations of motion read

$$W^{(1)}_{k,\boldsymbol{A}} = \begin{pmatrix} \Psi_{k,a}(\tau) \\ 0 \end{pmatrix} \,, \qquad (18)$$

where the expectation value of response fields by construction evaluates to zero.[10]

Connected two-point correlation functions are obtained from second-order functional derivatives and are given on the equations of motion by

$$W^{(2)}_{k,\boldsymbol{AB}} = \begin{pmatrix} C_{k,ab}(\tau,\tau',\Delta\boldsymbol{x}) & \mathrm{i}\, G^{\mathrm{R}}_{k,ab}(\tau,\tau',\Delta\boldsymbol{x}) \\ \mathrm{i}\, G^{\mathrm{A}}_{k,ab}(\tau,\tau',\Delta\boldsymbol{x}) & 0 \end{pmatrix} \,, \qquad (19)$$

where $\Delta\boldsymbol{x} = \boldsymbol{x} - \boldsymbol{x}'$ is the difference in position of the two fields. Here, $C_{k,ab}(\tau,\tau',\Delta\boldsymbol{x})$ is the covariance function of two physical fields and $G^{\mathrm{R}}_{k,ab}(\tau,\tau',\Delta\boldsymbol{x})$ is the (retarded) mean linear response function, in cosmology most often called propagator. The retarded propagator $G^{\mathrm{R}}_{k,ab}(\tau,\tau',\Delta\boldsymbol{x})$ is causal and vanishes for $\tau \le \tau'$.[11] Finally, the advanced and retarded propagators are related by

$$G^{\mathrm{A}}_{k,ab}(\tau,\tau',\boldsymbol{x}-\boldsymbol{x}') = G^{\mathrm{R}}_{k,ba}(\tau',\tau,\boldsymbol{x}'-\boldsymbol{x}) \,. \qquad (20)$$

Higher-order connected correlation functions can be obtained in a similar fashion from higher-order functional derivatives of $W_k$.

In the following, it is useful to introduce yet another generating functional, namely that of one-particle irreducible (1PI) correlation functions. The scale-dependent 1PI effective action is defined as the modified Legendre transform of $W_k$ with respect to the source currents $J_{\boldsymbol{A}}$,

$$\varGamma_k[\Psi; R] = \sup_J \left[ J_{\boldsymbol{A}} \Psi_{\boldsymbol{A}} - W_k \right] - \tfrac{1}{2}\, \Psi_{\boldsymbol{A}} R_{k,\boldsymbol{AB}}\, \Psi_{\boldsymbol{B}} \,. \qquad (21)$$

The term bilinear in the fields has been added for later convenience but vanishes in the absence of a regulator, in which case $\varGamma_k[\Psi; R] = \varGamma[\Psi]$ corresponds to the full 1PI effective action. It can be viewed as an analogue of the motion (8) in some time interval $\tau_{\mathrm{in}}$ to $\tau_{\mathrm{fi}}$ and depends on the method of discretisation for stochastic differential equations. To this end, the prescription equivalent to Itô calculus is employed, which is particular convenient since it does not need the introduction of additional ghost fields [45]. Integrals over the whole time interval are abbreviated as

$$\int_\tau = \int_{\tau_{\mathrm{in}}}^{\tau_{\mathrm{fi}}} \mathrm{d}\tau \,.$$

[9] When specifying whether the derivative is taken with respect to the physical or response field, the notation $Z^{(m,n)}_{A_1 \dots B_n}$ is used. Here, the first superscript corresponds to physical field derivatives, while the second superscript denotes response field derivatives.

[10] More generally, all expectation values $W^{(0,n)}_{B_1 \dots B_n}$ evaluate to zero at vanishing source currents due to general properties of the Janssen–de Dominicis formalism [46, 47].

[11] More rigorously, $G^{\mathrm{R}}_{k,ab}(\tau,\tau',\Delta\boldsymbol{x}) \to \delta_{ab}\theta(0)$ for $\tau' \to \tau^-$ which vanishes in the Itô prescription since it assigns $\theta(0) = 0$ to the Heaviside unit step function.

bare action (15) that already fully encodes all statistical information.[12]

By assuming that the right-hand side of equation (21) is maximised by some (field-dependent) source currents $J_{k,\boldsymbol{A}}[\Psi]$, one obtains the effective equations of motion

$$\Gamma^{(1)}_{k,\boldsymbol{A}} + R_{k,\boldsymbol{AB}}\,\Psi_{\boldsymbol{B}} = J_{k,\boldsymbol{A}}\ , \tag{22}$$

which determine the dynamics of the mean field. As remarked before, on the equations of motion only the physical mean field has non-trivial dynamics while the response mean field is vanishing.

By construction the connected and 1PI two-point correlation functions are inverse,

$$\left[\Gamma^{(2)}_{k,\boldsymbol{AB}} + R_{k,\boldsymbol{AB}}\right] W^{(2)}_{k,\boldsymbol{BC}} = \delta_{\boldsymbol{AC}}\ , \tag{23}$$

so that on the equations of motion one defines

$$\Gamma^{(2)}_{k,\boldsymbol{AB}} + R_{k,\boldsymbol{AB}} = \begin{pmatrix} 0 & -\mathrm{i}\,D^{\mathrm{A}}_{k,ab}(\tau,\tau',\Delta\boldsymbol{x}) \\ -\mathrm{i}\,D^{\mathrm{R}}_{k,ab}(\tau,\tau',\Delta\boldsymbol{x}) & H_{k,ab}(\tau,\tau',\Delta\boldsymbol{x}) \end{pmatrix}\ . \tag{24}$$

The inverse retarded propagator $D^{\mathrm{R}}_{k,ab}(\tau,\tau',\Delta\boldsymbol{x})$ is defined by the relation

$$\int_{\tau',\boldsymbol{x}'} D^{\mathrm{R}}_{k,ab}(\tau,\tau',\boldsymbol{x}-\boldsymbol{x}')\,G^{\mathrm{R}}_{k,bc}(\tau',\tau'',\boldsymbol{x}'-\boldsymbol{x}'') = \delta_{ac}\,\delta(\tau-\tau'')\,\delta(\boldsymbol{x}-\boldsymbol{x}'')\ , \tag{25}$$

and the 1PI statistical two-point correlation function $H_{k,ab}(\tau,\tau',\Delta\boldsymbol{x})$ is defined by

$$\int_{\tau',\boldsymbol{x}'} D^{\mathrm{R}}_{k,a\bar{a}}(\tau,\tau',\boldsymbol{x}-\boldsymbol{x}')\,C_{k,\bar{a}b}(\tau',\tau'',\boldsymbol{x}'-\boldsymbol{x}'') = \int_{\tau',\boldsymbol{x}'} H_{k,a\bar{b}}(\tau,\tau',\boldsymbol{x}-\boldsymbol{x}')\,G^{\mathrm{A}}_{k,\bar{b}b}(\tau',\tau'',\boldsymbol{x}'-\boldsymbol{x}'')\ . \tag{26}$$

The scale-dependent effective action is subject to the flow equation [17, 19]

$$\partial_k \Gamma_k = \tfrac{1}{2}\,\mathrm{Tr}\left[\left[\Gamma^{(2)}_k + R_k\right]^{-1}\cdot\partial_k R_k\right], \tag{27}$$

where the trace, dot and inverse operator are understood to run over time, space and the (response) field content.

The flow equation (27) is an exact functional differential equation and in the present context the analogue of Wetterich's equation [48]. Although the flow equation (27) can usually not be solved exactly, it is a very useful starting point for various (non-perturbative) investigations.

By applying functional derivatives to the flow equation (27) one obtains the flow of the one-point function

$$\partial_k \Gamma^{(1)}_{k,\boldsymbol{A}} = -\tfrac{1}{2}\,\mathrm{Tr}\left[W^{(2)}_k\cdot\Gamma^{(3)}_{k,\boldsymbol{A}}\cdot W^{(2)}_k\cdot\partial_k R_k\right]\ , \tag{28}$$

and two-point function,

$$\begin{aligned}\partial_k \Gamma^{(2)}_{k,\boldsymbol{AB}} = {}& \tfrac{1}{2}\,\mathrm{Tr}\left[W^{(2)}_k\cdot\Gamma^{(3)}_{k,\boldsymbol{A}}\cdot W^{(2)}_k\cdot\Gamma^{(3)}_{k,\boldsymbol{B}}\cdot W^{(2)}_k\cdot\partial_k R_k\right] \\ &+ (\boldsymbol{A}\longleftrightarrow\boldsymbol{B}) \\ &- \tfrac{1}{2}\,\mathrm{Tr}\left[W^{(2)}_k\cdot\Gamma^{(4)}_{k,\boldsymbol{AB}}\cdot W^{(2)}_k\cdot\partial_k R_k\right]\ .\end{aligned} \tag{29}$$

Here, relation (23) was used to replace the inverse of the effective action's second field derivatives with $W^{(2)}_k$ and should be understood as depending on the mean fields via the Legendre transformation (21).

The choice of a sensible regulator heavily depends on the problem at hand and the behaviour of the system in the infrared and ultraviolet. In the context of cosmology, where corrections to the bare action arise due to initial state fluctuations, the question of a sensible regulator is related to the scaling of the initial power spectrum. For a power law dark matter density contrast power spectrum, $P^{\mathrm{in}}_{\delta\delta}(q)\propto q^n$, corrections are finite in the infrared for $n>-1$ and in the ultraviolet for $n<-3$ to all orders in standard perturbation theory [49–52].[13] Realistic power spectra of the $\Lambda$CDM concordance model avoid divergences in both limits with a scaling $\propto q^{n_{\mathrm{s}}}$ in the infrared and $\propto q^{n_{\mathrm{s}}-4}\ln(q)^2$ in the ultraviolet with a pivot scale $q_*\approx 0.05\,\mathrm{Mpc}^{-1}$ and a (scalar) spectral index $n_{\mathrm{s}}\approx 0.96$ [53].

Since the theory is perturbative in the infrared, it is convenient to only regulate the ultraviolet part of the theory. To this end a regulator that only regulates the initial power spectrum is employed. In the following

$$R_{k,ab}(\tau,\tau',\boldsymbol{q}) = \delta(\tau-\tau_{\mathrm{in}})\,\delta(\tau'-\tau_{\mathrm{in}}) \times\left[P^{\mathrm{in}}_{k,ab}(\boldsymbol{q}) - P^{\mathrm{in}}_{ab}(\boldsymbol{q})\right]\begin{pmatrix}0 & 0\\ 0 & 1\end{pmatrix}\ , \tag{30}$$

is used and the scale-dependent initial power spectrum is chosen to be cut-off sharply for wave vectors above the renormalisation group scale, $P^{\mathrm{in}}_{k,ab}(\boldsymbol{q}) = \theta(k-q)\,P^{\mathrm{in}}_{ab}(\boldsymbol{q})$.[14]

---

[12] From a probabilistic point of view $Z_k$ and $W_k$ are moment- and cumulant-generating functionals of the 'probability density functional'

$$P_k[\psi] = \int \mathcal{D}\hat{\psi}\ \mathrm{e}^{-S_k[\psi,\hat{\psi}]+\hat{J}_A\hat{\psi}_A}\ ,$$

at least for a purely imaginary response field source current (corresponding to a real-valued source in the equations of motion of the physical fields). In the limit of a vanishing regulator the full 1PI effective action $\Gamma$ is related to a rate function which quantifies fluctuations away from the expected (mean field) behaviour, decaying asymptotically with $\exp(-\Gamma)$ for an infinite sample, at least in the standard ergodic paradigm.

[13] Even for finite corrections at all orders the perturbative series is only asymptotic and therefore does not need to converge [23].

[14] It has been criticised that for the regulator (30) the flow equation (27) simply describes initial power spectrum variations rather than truly capturing the effects of coarse-gaining [54]. While it is true that modes which are initially absent can (and will) be dynamically generated, the flow equation (27) is not restricted to these types of regulators. In principle, the dynamical part of the bare action (15) can also be regulated such that the propagation of modes on scales $q>k$ is essentially absent.

In the limit $k \to 0$ the initial power spectrum $P^{\text{in}}_{k,ab}(\boldsymbol{q})$ vanishes so that no initial state fluctuations are included. In this limit the scale-dependent effective action equals the bare action (15), i.e.

$$\lim_{k\to 0} \Gamma_k = S\,. \tag{31}$$

In the opposite limit, $k \to \infty$, all fluctuations are included and as such the scale-dependent effective action equals the full effective action,

$$\lim_{k\to \infty} \Gamma_k = \Gamma\,. \tag{32}$$

Using the limit (31) as an initial condition, the flow equation (27) can be used to find $\Gamma_k$ at any scale $k$ and in particular also in the limit of equation (32). This provides a non-perturbative possibility to study the influence of initial state fluctuations for non-linear cosmological structure formation.

Having specified the regulator, the flow equations (28) and (29) can be given more explicitly. Diagrammatically, the flow of the effective equations of motion reads

$\partial_k \Gamma^{(0,1)}_{k,A} = -\frac{1}{2}$ 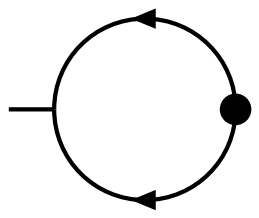 , (33)

while the inverse propagator flow is given by

$\partial_k \Gamma^{(1,1)}_{k,AB} = \frac{1}{2}$ 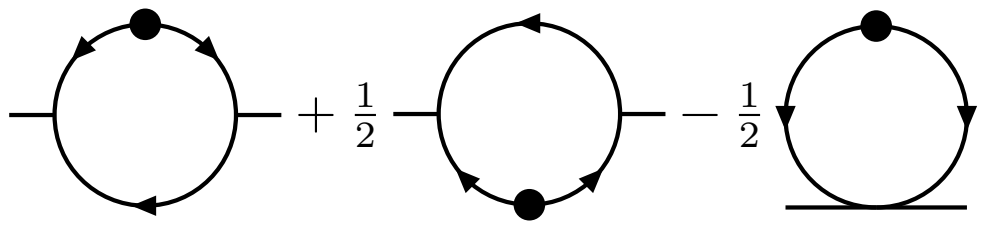 , (34)

and the statistical two-point function flow reads

$\partial_k \Gamma^{(0,2)}_{k,AB} = \frac{1}{2}$ 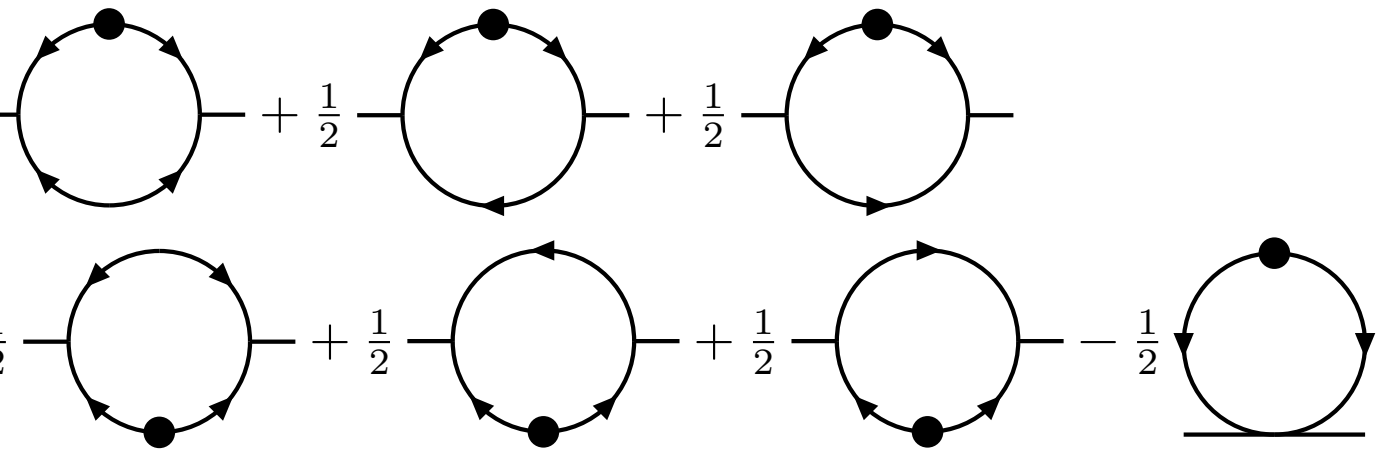 . (35)

Here the black dot denotes the regulated initial power spectrum, an edge with a single arrowhead is a propagator, while an edge with two (opposite) arrowheads corresponds to a power spectrum. The arrowheads indicate the direction of increasing time.

## IV. SYMMETRIES AND RELATED WARD IDENTITIES

In this section symmetries of the bare action (15) are investigated in order to understand the general structure of the effective action $\Gamma_k$ and derive related generalised Ward identities [55]. The studied symmetries correspond to (infinitesimal) affine field transformations

$$\delta_\epsilon \psi_{\boldsymbol{A}} = L_{\epsilon,\boldsymbol{AB}}\, \psi_{\boldsymbol{A}} + T_{\epsilon,\boldsymbol{A}}\,, \tag{36}$$

where $L_\epsilon$ is a linear operator and $T_\epsilon$ a translation in field space. In this context, a transformation that leaves the action invariant is called a true symmetry, while an extended symmetry changes the action by terms that are at most linear in the fields [56, 57]. Since a change of integration variables must leave the generating functional (14) unaltered, one obtains the Ward identity

$$\Gamma^{(1)}_k \cdot \delta_\epsilon \Psi = \delta_\epsilon S[\Psi] + \mathrm{Tr}\Big[L_\epsilon \cdot W^{(2)}_k \cdot R_k\Big]\,. \tag{37}$$

The first term on the right-hand side only contributes for extended symmetries, while the second term vanishes if the regulator respects the symmetry transformation.[15] Since the employed regulator (30) only alters the initial power spectrum, the regulated action (17) respects the same symmetries as the bare action (15) so that the second term on the right-hand side of the Ward identity (37) vanishes.

### A. Conservation of mass

Conservation of mass is ensured at the level of the bare action (15) by the continuity equation. This extends to the effective action and is related to a time-gauged density contrast response field shift, $\delta_\epsilon\hat{\delta}(\tau,\boldsymbol{x}) = \epsilon(\tau)$, which changes the bare action by a term linear in the fields,

$$\begin{aligned}\delta_\epsilon S = &-\mathrm{i}\int_{\tau,\boldsymbol{x}} \epsilon(\tau)\,\partial_\tau\delta(\tau,\boldsymbol{x}) \\ &+ \int_{\boldsymbol{x},\boldsymbol{x}'} \epsilon(\tau_\mathrm{in})\,C^\mathrm{in}_{\delta a}(\boldsymbol{x}-\boldsymbol{x}')\,\hat{\varPsi}_a(\tau_\mathrm{in},\boldsymbol{x}')\ ,\end{aligned} \tag{38}$$

where the second term on the right-hand side vanishes for the type of initial power spectra considered here since it is proportional to $P^\mathrm{in}_{ab}(\boldsymbol{0})$.[16] Since $\epsilon(\tau)$ is an (infinitesimal) arbitrary function of time, one obtains the Ward identity

$$\int_{\boldsymbol{x}} \varGamma^{(0,1)}_{k,\delta}(\tau,\boldsymbol{x}) = -\mathrm{i}\int_{\boldsymbol{x}} \partial_\tau\delta(\tau,\boldsymbol{x})\ , \tag{39}$$

which encodes that the effective equations of motion of the density contrast field are of conservative form. The Ward identity may equivalently be written in Fourier space as

$$\varGamma^{(0,1)}_{k,\delta}(\tau,\boldsymbol{0}) = -\mathrm{i}\,\partial_\tau\delta(\tau,\boldsymbol{0})\ . \tag{40}$$

[15] Additionally, it is assumed that the functional integral measure is invariant under the symmetry transformation, i.e. in the absence of an anomaly, which is the case for the two symmetries studied in this section.

Here and in the following, specific mean fields are denoted by the same symbols as their fluctuation counterpart for a clearer notation.

### B. Extended Galilean invariance

A symmetry of the late-time Universe is Galilean invariance and naturally also holds for the Vlasov–Poisson equations [51, 58–60]. In an expanding spacetime a Galilean transformation corresponds to the (time-dependent) comoving coordinate and conjugate momentum change

$$\boldsymbol{x} \mapsto \boldsymbol{x} + \boldsymbol{v}T\ , \qquad \boldsymbol{p} \mapsto \boldsymbol{p} + am\boldsymbol{v}\dot{T}\ , \tag{41}$$

where $\boldsymbol{v}$ is a constant velocity and

$$T(\tau) = \frac{1}{a(\tau)}\int_{\tau_\mathrm{in}}^{\tau} \mathrm{d}\tau'\,a(\tau')\ . \tag{42}$$

Under this coordinate change the cumulants of the distribution function transform as

$$c^{(n)}_{i_1\dots i_n}(\tau,\boldsymbol{x}) \mapsto c^{(n)}_{i_1\dots i_n}(\tau,\boldsymbol{x}-\boldsymbol{v}T) + \delta_{n1}v_{i_1}\dot{T}\ , \tag{43}$$

so that only the velocity field is non-trivially shifted. The equations of motion (3) are invariant under this transformation up to a time-dependent shift in the velocity field equation. While this shift is compensated for by the transformation of the gravitational potential [51]

$$\partial_i\phi(\tau,\boldsymbol{x}) \mapsto \partial_i\phi(\tau,\boldsymbol{x}-\boldsymbol{v}T) + v_iT\dot{\mathcal{H}}\ , \tag{44}$$

the symmetry is no longer apparent when the gravitational potential is eliminated by solving Poisson's equation. Indeed, Galilean invariance is no longer manifest in the formulation (8) since integrating Poisson's equation in terms of the operator (10) fixes a frame with respect to which expectation values are computed.[17] In this sense the Galilean transformation (43) is already an extended symmetry that changes the bare action (15) by terms linear in the velocity response field.

The transformation (43) extends to a time-gauged symmetry of the effective action for the infinitesimal field transformations

$$\begin{aligned}\delta_\epsilon\psi_a(\tau,\boldsymbol{x}) &= -\epsilon_i(\tau)\,\partial_i\psi_a(\tau,\boldsymbol{x}) + \delta_{au_i}\dot{\epsilon}_i(\tau)\ ,\\ \delta_\epsilon\hat{\psi}_a(\tau,\boldsymbol{x}) &= -\epsilon_i(\tau)\,\partial_i\hat{\psi}_a(\tau,\boldsymbol{x})\ .\end{aligned} \tag{45}$$

Under these transformations all terms in the bare action (15) are invariant except for the term involving the time derivative and the Hubble drag term of the velocity field, giving rise to

$$\delta_\epsilon S = -\mathrm{i}\int_{\tau,\boldsymbol{x}} \hat{u}_i(\tau,\boldsymbol{x})\Big[\ddot{\epsilon}_i(\tau) + \mathcal{H}\dot{\epsilon}_i(\tau)\Big]\ . \tag{46}$$

Since the right-hand side is linear in fields, this corresponds to an extended symmetry [56, 57]. The corresponding Ward identity reads

$$\begin{aligned}&\int_{\boldsymbol{x}} \Big[\varPsi_a(\tau,\boldsymbol{x})\,\partial_i - \delta_{au_i}\,\partial_\tau\Big]\varGamma^{(1,0)}_{k,a}(\tau,\boldsymbol{x}) \\ &\qquad + \int_{\boldsymbol{x}} \hat{\varPsi}_a(\tau,\boldsymbol{x})\,\partial_i\varGamma^{(0,1)}_{k,a}(\tau,\boldsymbol{x}) \\ &\qquad = -\mathrm{i}\int_{\boldsymbol{x}} \Big[\partial_\tau^2 - \mathcal{H}\partial_\tau - \dot{\mathcal{H}}\Big]\hat{u}_i(\tau,\boldsymbol{x})\ ,\end{aligned} \tag{47}$$

and entails that, apart from the velocity field's time derivative and Hubble drag term which are not renormalised, the effective action is invariant under time-gauged Galilean transformations.

Applying field derivatives to the Ward identity (47) yields related identities such that for $m+n>1$ one obtains in Fourier space

$$\begin{aligned}&\left[\sum_{l=1}^{m}\theta(\tau-\tau_l)\,\mathrm{i}\,q_{l,i} + \sum_{\bar{l}=1}^{n}\theta(\tau-\tau'_{\bar{l}})\,\mathrm{i}\,q'_{\bar{l},i}\right] \\ &\qquad\times\varGamma^{(m,n)}_{k,a_1\dots b_n}(\tau_1,\boldsymbol{q}_1;\dots;\tau'_n,\boldsymbol{q}'_n) \\ &\qquad = \varGamma^{(m+1,n)}_{k,u_ia_1\dots b_n}(\tau,\boldsymbol{0};\tau_1,\boldsymbol{q}_1;\dots;\tau'_n,\boldsymbol{q}'_n)\ .\end{aligned} \tag{48}$$

[16] Typically, suitable decay or periodic boundary conditions are imposed in the functional integral so that the space integral over the total derivative term in the continuity equation vanishes.

[17] Equivalently, one can keep the gravitational potential and Poisson's equation at the expense of introducing another response field so that Galilean invariance is manifest as shown in appendix A.

The Ward identities (48) impose linear relations between 1PI correlation functions of order $(m+1, n)$ at vanishing wave vector for a velocity field and 1PI correlation functions of lower order $(m, n)$.

## V. LARGE EXTERNAL WAVE NUMBER LIMIT AND THE SWEEPING EFFECT

In this section the large external wave number limit of the 1PI two-point correlation function's flow equations is investigated. Using the Ward identities (48) related to extended Galilean invariance, the flow equations can be (formally) closed in this limit, at least in the absence of higher-order velocity cumulants. The procedure presented here is closely related to the large external wave number limit studied in the context of fluid turbulence [61–63], although being in a non-stationary setting in cosmology.

### A. Large external wave number limit

In order to derive the large external wave number limit of the flow equations (34) and (35), the first diagram of the inverse propagator flow (34) is considered as an illustrative example. It is given by

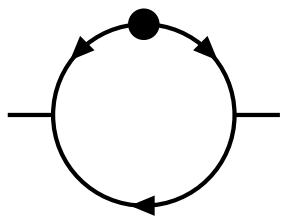

$$= -\mathrm{i}\,\delta(\boldsymbol{q}+\boldsymbol{q}')\int_{\substack{\bar\tau,\bar\tau'\\ \tilde\tau,\tilde\tau'}}\int_{\boldsymbol{l}}\hat\Gamma^{(2,1)}_{k,cea}(\tilde\tau,\boldsymbol{l};\bar\tau,-\boldsymbol{q}-\boldsymbol{l};\tau,\boldsymbol{q})\,G^{\mathrm{R}}_{k,ef}(\bar\tau,\bar\tau',\boldsymbol{q}+\boldsymbol{l})\,\hat\Gamma^{(2,1)}_{k,dbf}(\tilde\tau',-\boldsymbol{l};\tau',-\boldsymbol{q};\bar\tau',\boldsymbol{q}+\boldsymbol{l}) \\ \times\hat\partial_k P^{\mathrm{I}}_{k,dc}(\tilde\tau',\tilde\tau,\boldsymbol{l})\,, \tag{49}$$

where the circumflex denotes that an overall wave vector conserving delta function has been extracted from the 1PI three-point functions. Further, the abbreviation

$$P^{\mathrm{I}}_{k,ab}(\tau,\tau',\boldsymbol{q}) = G^{\mathrm{R}}_{k,a\bar a}(\tau,\tau_{\mathrm{in}},\boldsymbol{q})\,P^{\mathrm{in}}_{k,\bar a\bar b}(\boldsymbol{q}) \\ \times G^{\mathrm{A}}_{k,\bar b b}(\tau_{\mathrm{in}},\tau',\boldsymbol{q})\,, \tag{50}$$

is used and the derivative

$$\hat\partial_k = \int_{\boldsymbol{q}}\partial_k P^{\mathrm{in}}_{k,ab}(\boldsymbol{q})\,\frac{\delta}{\delta P^{\mathrm{in}}_{k,ab}(\boldsymbol{q})}\,, \tag{51}$$

only acts on the regulated initial power spectrum.

The internal wave vector $\boldsymbol{l}$ running through the loop of the diagram (49) is restricted to $|\boldsymbol{l}| = k$ due to the regulator (30). In the limit $q\to\infty$ the internal wave vector $\boldsymbol{l}$ is therefore small in magnitude compared to $\boldsymbol{q}$ and may be set to zero within the 1PI three-point functions.[18] In the case where the vanishing wave vector is assigned to a velocity mode the Ward identity (48) can be used to relate the 1PI three-point function to a 1PI two-point function.

*A priori* it is not clear why the vanishing wave vector should be assigned to velocity modes since the loop naturally runs over all degrees of freedom included in the field content (7). In the following, it is argued that in the absence of velocity dispersion and higher-order velocity cumulants it is expected that in the limit $q\to\infty$ the leading contribution is due to the velocity-velocity sector of the regulator. More specifically, it is shown that in the large external wave number limit the diagram (49) is given by

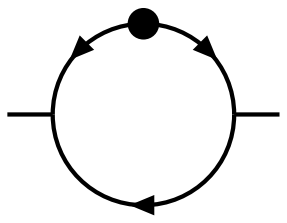

$$\sim -\mathrm{i}\,\delta(\boldsymbol{q}+\boldsymbol{q}')\int_{\substack{\bar\tau,\bar\tau'\\ \tilde\tau,\tilde\tau'}}\hat\Gamma^{(2,1)}_{k,u_iea}(\tilde\tau,\boldsymbol{0};\bar\tau,-\boldsymbol{q};\tau,\boldsymbol{q})\,G^{\mathrm{R}}_{k,ef}(\bar\tau,\bar\tau',\boldsymbol{q})\,\hat\Gamma^{(2,1)}_{k,u_jbf}(\tilde\tau',\boldsymbol{0};\tau',-\boldsymbol{q};\bar\tau',\boldsymbol{q}) \\ \times\hat\partial_k\int_{\boldsymbol{l}}P^{\mathrm{I}}_{k,u_ju_i}(\tilde\tau',\tilde\tau,\boldsymbol{l})\,, \tag{52}$$

[18] Strictly speaking, this is only possible if the correlation functions are analytic in wave vectors. The non-gradient dependence due to the operator (10) implies non-analyticity of the (inverse) propagator in the velocity-density component. The corresponding infrared divergence is associated with homogeneous mass density shifts and is usually treated by regularising gravitational interactions at large scales and related to the *Jeans swindle* [10, 64]. In the following, the tacit assumption is made that no other non-analyticities develop in the presence of a regulator.

at least *perturbatively* to all orders. The line of argument presented here is very similar to the classification of diagrams in renormalised perturbation theory [2, 3].

Consider the diagram (49) and amputate the regulator $\hat{\partial}_k P^{\mathrm{I}}_{k,cd}$. At lowest order in standard perturbation theory the leading contribution in the limit $q \to \infty$ is given by

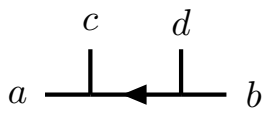

$$\sim\ q^2\, g^{\mathrm{R}}_{ab}(\tau,\tau',\boldsymbol{q})\ \delta_{cu_i}\delta_{du_j} \times (q\text{-ind.})\ , \quad (53)$$

where the edge now denotes a linear propagator and the vertices are bare. The limit makes use of the fact that the wave vector $\boldsymbol{l}$ is bounded in magnitude due to the regulator and thus negligible compared to $\boldsymbol{q}$. The leading contribution is then due to the scaling of the linear propagator

$$g^{\mathrm{R}}_{ab}(\tau,\tau',\boldsymbol{q}) \sim \begin{pmatrix} 1 & q_j \\ q_i/q^2 & \delta_{ij} \end{pmatrix} \times (q\text{-ind.})\ , \quad (54)$$

and the structure of the bare vertices.

At the next higher order in perturbation theory two types of diagrams need to be distinguished. In the language of renormalised perturbation theory one can realise that every perturbative diagram has a *principle path* that connects the ingoing and outgoing mode with a chain of linear retarded propagators. Diagrams can be organised according to how many interactions (via bare vertices) are along this path. Since for each bare vertex that is passed along the principal path a factor $q$ is picked up in the large external wave number limit, the leading contribution is due to diagrams where all interactions are on that path.

As an example consider the contributions where one vertex of the diagram (53) is evaluated at one-loop. These consist of diagrams of the type

$\mathrm{D}_1 =$

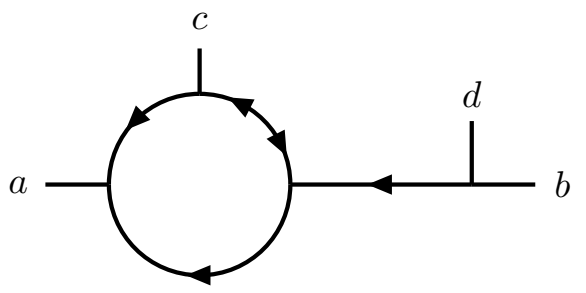

, (55)

and

$\mathrm{D}_2 =$

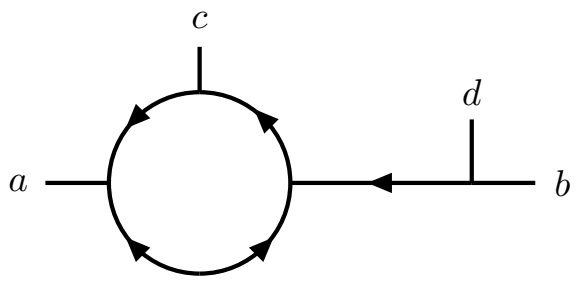

. (56)

Similarly, evaluating the retarded propagator in diagram (53) at one-loop one obtains

$\mathrm{D}_3 =$

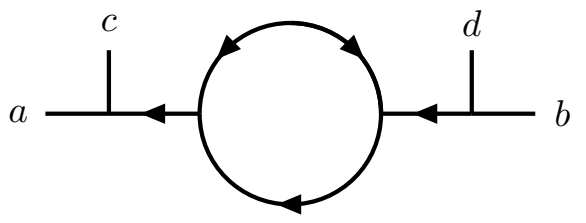

. (57)

The leading contribution of the diagrams is obtained by counting the vertices along the principal path such that one obtains

$$\begin{aligned} \mathrm{D}_1 &\sim q^3\, g^{\mathrm{R}}_{ab}(\tau,\tau',\boldsymbol{q})\ \delta_{du_i} \times (q\text{-ind.})\ , \\ \mathrm{D}_{2,3} &\sim q^4\, g^{\mathrm{R}}_{ab}(\tau,\tau',\boldsymbol{q})\ \delta_{cu_i}\delta_{du_j} \times (q\text{-ind.})\ . \end{aligned} \quad (58)$$

This argument extends to any perturbative order and can be applied to all diagrams entering the flow equations (34) and (35). In turn only the velocity-velocity part of the regulator $\hat{\partial}_k P^{\mathrm{I}}_{k,u_j u_i}$ enters into the expression (52), at least to leading order.

Although the presented argument holds to all orders in perturbation theory, there is no rigorous justification why it should hold non-perturbatively. More specifically, in the presence of non-perturbative scales the line of argument presented here can not be straight forwardly extended to full propagators and vertices.

Further, the presented argument no longer holds in the presence of higher-order velocity cumuants, e.g. velocity dispersion. Indeed, considering again the lowest order contribution (53) in the presence of velocity dispersion and taking into account the allowed vertices due to the non-linear terms in equation (8), one finds the leading contribution to the first diagram of the velocity-velocity inverse propagator flow to be

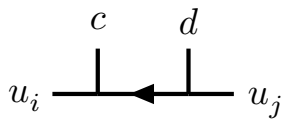

$$\sim\ q^2\, g^{\mathrm{R}}_{\delta\sigma}(\tau,\tau',\boldsymbol{q})\ \delta_{c\sigma}\delta_{d\sigma} \times (q\text{-ind.})\ , \quad (59)$$

where tensorial structures are suppressed on the right-hand side. This contribution dominates in the limit $q \to \infty$ over the one given in equation (53) due to the scaling

$$g^{\mathrm{R}}_{\delta\sigma_{ij}}(\tau,\tau',\boldsymbol{q}) \sim q_i q_j \times (q\text{-ind.})\ . \quad (60)$$

The justification that the vanishing wave vector is assigned to a velocity mode therefore relies on the following two assumptions:

- The emergence of non-perturbative scales does not invalidate the classification of leading contributions described before.
- Higher-order velocity cumulants are absent or at least subdominant.

Under these assumptions the limit (52) holds and the 1PI three-point vertices can be replaced using the Ward identity (48). This can be done for all diagrams entering the flow equations (34) and (35). More precisely, the three-point functions are replaced by

$$\hat{\Gamma}^{(2,1)}_{k,u_iba}(\tau'',\mathbf{0};\tau',-\boldsymbol{q};\tau,\boldsymbol{q}) = -\mathrm{i}\,q_i\big[\theta(\tau''-\tau')-\theta(\tau''-\tau)\big]\hat{\Gamma}^{(1,1)}_{k,ba}(\tau',-\boldsymbol{q};\tau,\boldsymbol{q})\,,$$
$$\hat{\Gamma}^{(1,2)}_{k,u_iba}(\tau'',\mathbf{0};\tau',-\boldsymbol{q};\tau,\boldsymbol{q}) = -\mathrm{i}\,q_i\big[\theta(\tau''-\tau')-\theta(\tau''-\tau)\big]\hat{\Gamma}^{(0,2)}_{k,ba}(\tau',-\boldsymbol{q};\tau,\boldsymbol{q})\,, \tag{61}$$

whereas the four-point functions are replaced using

$$\hat{\Gamma}^{(3,1)}_{k,u_iu_jba}(\tau''',\mathbf{0};\tau'',\mathbf{0};\tau',-\boldsymbol{q};\tau,\boldsymbol{q}) = -q_iq_j\big[\theta(\tau'''-\tau')-\theta(\tau'''-\tau)\big]\big[\theta(\tau''-\tau')-\theta(\tau''-\tau)\big]\hat{\Gamma}^{(1,1)}_{k,ba}(\tau',-\boldsymbol{q};\tau,\boldsymbol{q})\,,$$
$$\hat{\Gamma}^{(2,2)}_{k,u_iu_jba}(\tau''',\mathbf{0};\tau'',\mathbf{0};\tau',-\boldsymbol{q};\tau,\boldsymbol{q}) = -q_iq_j\big[\theta(\tau'''-\tau')-\theta(\tau'''-\tau)\big]\big[\theta(\tau''-\tau')-\theta(\tau''-\tau)\big]\hat{\Gamma}^{(0,2)}_{k,ba}(\tau',-\boldsymbol{q};\tau,\boldsymbol{q})\,. \tag{62}$$

Substituting the relations (61) and (62) into the inverse propagator flow (34) one obtains

$$\partial_k D^{\mathrm{R}}_{k,ab}(\tau,\tau',\boldsymbol{q}) = q^2\int_{\bar\tau,\bar\tau'} D^{\mathrm{R}}_{k,ae}(\tau,\bar\tau,\boldsymbol{q})\,G^{\mathrm{R}}_{k,ef}(\bar\tau,\bar\tau',\boldsymbol{q})\,D^{\mathrm{R}}_{k,fb}(\bar\tau',\tau',\boldsymbol{q})\,\hat\partial_k I_k(\tau,\bar\tau;\bar\tau',\tau') - \frac{q^2}{2}\,D^{\mathrm{R}}_{k,ab}(\tau,\tau',\boldsymbol{q})\,\hat\partial_k I_k(\tau,\tau';\tau,\tau')\,, \tag{63}$$

and similarly for the statistical two-point function flow (35)

$$\begin{aligned}\partial_k H_{k,ab}(\tau,\tau',\boldsymbol{q}) = q^2\int_{\bar\tau,\bar\tau'}\Big[&D^{\mathrm{R}}_{k,ae}(\tau,\bar\tau,\boldsymbol{q})\,P_{k,ef}(\bar\tau,\bar\tau',\boldsymbol{q})\,D^{\mathrm{A}}_{k,fb}(\bar\tau',\tau',\boldsymbol{q})\\ &- D^{\mathrm{R}}_{k,ae}(\tau,\bar\tau,\boldsymbol{q})\,G^{\mathrm{R}}_{k,ef}(\bar\tau,\bar\tau',\boldsymbol{q})\,H_{k,fb}(\bar\tau',\tau',\boldsymbol{q})\\ &- H_{k,ae}(\tau,\bar\tau,\boldsymbol{q})\,G^{\mathrm{A}}_{k,ef}(\bar\tau,\bar\tau',\boldsymbol{q})\,D^{\mathrm{A}}_{k,fb}(\bar\tau',\tau',\boldsymbol{q})\Big]\hat\partial_k I_k(\tau,\bar\tau;\tau',\bar\tau')\\ &- \frac{q^2}{2}\,H_{k,ab}(\tau,\tau',\boldsymbol{q})\,\hat\partial_k I_k(\tau,\tau';\tau,\tau') + \delta(\tau-\tau_{\mathrm{in}})\,\delta(\tau'-\tau_{\mathrm{in}})\,\hat\partial_k P^{\mathrm{in}}_{k,ab}(\boldsymbol{q})\,,\end{aligned} \tag{64}$$

where the function $I_k$ is given by

$$I_k(\tau,\tau';\bar\tau,\bar\tau') = \frac{1}{3}\int_{\tau'}^{\tau}\mathrm{d}\tau''\int_{\bar\tau'}^{\bar\tau}\mathrm{d}\bar\tau''\int_{\boldsymbol{q}} P^{\mathrm{I}}_{k,u_iu_i}(\tau'',\bar\tau'',\boldsymbol{q})\,. \tag{65}$$

As a concrete example the first diagram of the inverse propagator flow (34) is computed in appendix B and the other diagrams follow similarly.

The flow equations (63) and (64) are (formally) closed at the level of two-point functions, although involving connected and 1PI correlation functions. The flow of the propagator and power spectrum can be obtained from relation (23) and read

$$\partial_k G^{\mathrm{R}}_{k,ab}(\tau,\tau',\boldsymbol{q}) = -\int_{\bar\tau,\bar\tau'} G^{\mathrm{R}}_{k,a\bar a}(\tau,\bar\tau,\boldsymbol{q})\times\partial_k D^{\mathrm{R}}_{k,\bar a\bar b}(\bar\tau,\bar\tau',\boldsymbol{q})\times G^{\mathrm{R}}_{k,\bar bb}(\bar\tau',\tau',\boldsymbol{q})\,, \tag{66}$$

and

$$\partial_k P_{k,ab}(\tau,\tau',\boldsymbol{q}) = \int_{\bar\tau,\bar\tau'}\partial_k\Big[G^{\mathrm{R}}_{k,a\bar a}(\tau,\bar\tau,\boldsymbol{q})\times H_{k,\bar a\bar b}(\bar\tau,\bar\tau',\boldsymbol{q})\times G^{\mathrm{A}}_{k,\bar bb}(\bar\tau',\tau',\boldsymbol{q})\Big]\,. \tag{67}$$

Using the flow equations (63) and (64) one finally arrives at the rather simple equation for the retarded propagator

$$\partial_k G^{\mathrm{R}}_{k,ab}(\tau,\tau',\boldsymbol{q}) = -\tfrac{1}{2}\,q^2\,\hat\partial_k I_k(\tau,\tau';\tau,\tau')\times G^{\mathrm{R}}_{k,ab}(\tau,\tau',\boldsymbol{q})\,, \tag{68}$$

and the power spectrum

$$\partial_k P_{k,ab}(\tau,\tau',\boldsymbol{q}) = -\tfrac{1}{2}\,q^2\,\hat\partial_k I_k(\tau,\tau';\tau,\tau')\times P_{k,ab}(\tau,\tau',\boldsymbol{q}) + \hat\partial_k P^{\mathrm{I}}_{k,ab}(\tau,\tau',\boldsymbol{q})\,. \tag{69}$$

Note that the function $I_k$ defined in equation (65) vanishes at equal times of either of the two time argument pairs due to the Heaviside unit step functions appearing in the identities (61) and (62) as well as being localised at the renormalisation group scale $k$ due to the regulator. This implies in particular that the first term on the right-hand side of equation (69) vanishes for the equal-time power spectrum.

Although the equations (63) and (64) are formally closed, the function $I_k$ involves knowledge of the propagator at $q = k$ which is the opposite limit to what was assumed in the derivation, at least in some regions of the renormalisation group flow trajectories.

While the function $I_k$ is non-universal, the fact that the propagator and power spectrum have a Gaussian suppression in wave number $q$ in the limit $q \to \infty$ is universal and a direct results of the possibility to close the flow equations at the level of two-point functions. As was remarked before, this holds as long as dark matter is described by the single-stream approximation in the absence of non-perturbative scales and other effects due to e.g. velocity dispersion are not present. In turn, any violation from this scaling has to be due to the emergence of non-perturbative scales or due to higher-order velocity cumulants and is regarded as an interesting possible signature for such non-perturbative effects.

### B. Sweeping effect

A simple approximation that allows to solve the flow equations (68) and (69) analytically is given by evaluating the propagators in the expression $\hat{\partial}_k P^{\mathrm{I}}_{k,ab}$ at linear level so that $\hat{\partial}_k P^{\mathrm{L}}_{k,ab}$ is the regulator entering the flow equations. This is justified for a renormalisation group flow deep in the infrared, where gravitational dynamics is well described by linear theory. There, one obtains

$$I_k(\tau,\tau';\tau,\tau') = \frac{[D_+(\tau)-D_+(\tau')]^2}{D_+(\tau_{\mathrm{in}})^2}\,\sigma^2_{\mathrm{v},k}\;, \tag{70}$$

for growing mode initial conditions,

$$u_i^{\mathrm{L}}(\tau,\boldsymbol{q}) = \frac{\mathrm{i}\,q_i}{q^2}\frac{\dot{D}_+(\tau)}{D_+(\tau)}\,\delta^{\mathrm{L}}(\tau,\boldsymbol{q})\;, \tag{71}$$

where $D_+$ is the standard linear growing mode of density fluctuations in the single-stream approximation, here normalised to unity at $a=1$, corresponding to today. Further, $\sigma_{\mathrm{v},k}$ is the initial root-mean-square velocity,

$$\begin{aligned}\sigma^2_{\mathrm{v},k} &= \frac{1}{3}\,C^{\mathrm{in}}_{k,u_iu_i}(\boldsymbol{0})\Big/\frac{\dot{D}_+(\tau_{\mathrm{in}})^2}{D_+(\tau_{\mathrm{in}})^2}\\ &= \frac{1}{6\pi^2}\int_0^k \mathrm{d}q\,q^2\,P^{\mathrm{in}}_{u_iu_i}(\boldsymbol{q})\Big/\frac{\dot{D}_+(\tau_{\mathrm{in}})^2}{D_+(\tau_{\mathrm{in}})^2}\;,\end{aligned} \tag{72}$$

up to a factor of $\dot{D}_+(\tau_{\mathrm{in}})^2/D_+(\tau_{\mathrm{in}})^2$. The flow equation for the propagator is then solved by

$$\begin{aligned}G^{\mathrm{R}}_{k,ab}(\tau,\tau',\boldsymbol{q}) &= g^{\mathrm{R}}_{ab}(\tau,\tau',\boldsymbol{q})\\ &\quad\times \mathrm{e}^{-\frac{1}{2}q^2\sigma^2_{\mathrm{v},k}[D_+(\tau)-D_+(\tau')]^2/D_+(\tau_{\mathrm{in}})^2}\;,\end{aligned} \tag{73}$$

and similar for the power spectrum,

$$\begin{aligned}P_{k,ab}(\tau,\tau',\boldsymbol{q}) &= P^{\mathrm{L}}_{k,ab}(\tau,\tau',\boldsymbol{q})\\ &\quad\times \mathrm{e}^{-\frac{1}{2}q^2\sigma^2_{\mathrm{v},k}[D_+(\tau)-D_+(\tau')]^2/D_+(\tau_{\mathrm{in}})^2}\;.\end{aligned} \tag{74}$$

In this setting the propagator and the unequal-time power spectrum feature a Gaussian suppression factor due to the *linear* root-mean-square velocity field.

In the following it is shown that a random background flow, associated to a velocity mean field, has the same effect on the linear response function and is related to the *sweeping effect* previously discussed in the context of fluid turbulence [65]. To this end consider the cumulant evolution equations (3) on a background flow $v_i(\tau)$. It is assumed that the background flow evolves proportional to some function $\dot{\mu}(\tau)$ so that $v_i(\tau)=\dot{\mu}(\tau)\,v_i$, where $v_i$ is a zero-mean normal distributed multivariate random variable.[19] The linear response function is the Green's function of the linear equations of motion (8) which are modified in the presence of a background flow to

$$\begin{aligned}\partial_\tau\psi_a(\tau,\boldsymbol{x}) + \int_{\boldsymbol{x}'} \Omega_{ab}(\tau,\boldsymbol{x}-\boldsymbol{x}')\,\psi_b(\tau,\boldsymbol{x}')&\\ + v_i(\tau)\,\partial_i\psi_a(\tau,\boldsymbol{x}) = 0\;.&\end{aligned} \tag{75}$$

The corresponding linear response function is then given in Fourier space by

$$g^{\mathrm{R}}_{ab}(\tau,\tau',\boldsymbol{q})\,\mathrm{e}^{-\mathrm{i}\boldsymbol{q}\cdot\boldsymbol{v}[\mu(\tau)-\mu(\tau')]}\;, \tag{76}$$

where $g^{\mathrm{R}}_{ab}(\tau,\tau',\boldsymbol{q})$ is the linear response in the absence of a background flow. The mean linear response function is then given by averaging over the distribution of $v_i$ such that one obtains [65]

$$G^{\mathrm{R}}_{ab}(\tau,\tau',\boldsymbol{q}) = g^{\mathrm{R}}_{ab}(\tau,\tau',\boldsymbol{q})\,\mathrm{e}^{-\frac{1}{2}q^2v^2_{\mathrm{rms}}[\mu(\tau)-\mu(\tau')]^2}\;, \tag{77}$$

where $v_{\mathrm{rms}}$ is the root-mean-square background velocity.

This analysis shows that the Gaussian suppression factor in the propagator (77) is not related to a true loss of memory due to relaxation processes but rather to the random advection of small-scale structures due to a large-scale flow also know as sweeping effect [11, 66–68].

One can now notice that the large wave number limit propagator (73) is of a similar form as the response function on a random background flow. In the infrared of the renormalisation group flow the suppression is due to the linear root-mean-square velocity, suggesting that it does not truly capture the effect of memory loss associated with relaxation towards equilibrium but rather describes the sweeping of small-scale structure due to an effective random large-scale advection. In contrast, the flow equations (68) and (69) are more general since the function $I_k$ includes non-linear information beyond the sweeping effect.

The propagator (73) was first obtained in the framework of renormalised perturbation theory [3]. Interestingly enough, this form of the propagator is actually exact in the Zeldovich approximation [66], whereas for more realistic approximations, such as the adhesion model, the propagator is already much more complicated [67, 68].

## VI. CONCLUSIONS

The paper discussed an approach to cosmic large-scale structure formation based on the non-perturbative functional renormalisation group. The basic idea is to modify the initial power spectrum of dark matter mass density fluctuations and to study how the 1PI effective action changes. This is a conceptually interesting application of the functional renormalisation group formalism because for any value of the renormalisation group scale $k$, which parametrises the modification of the initial power spectrum, one has a viable physical theory. An immediate benefit of this setup is that the regularisation breaks no symmetries. Further, one can choose the initial power spectrum such that it corresponds to dark matter with different thermal production temperatures, encompassing warm dark matter models where the power spectrum is concentrated at small wave numbers and very cold dark matter models where the power spectrum extends far into the ultraviolet regime.

[19] For a velocity field decaying with the Hubble expansion one simply has $\dot{\mu}(\tau)\propto a(\tau)^{-1}$.

One of the main topics investigated here were symmetries and related Ward identities. Particularly interesting and powerful is Galilean invariance and a time-gauged extension thereof. In a formalism where the gravitational potential is integrated out, such transformations should be seen as an extended symmetry which change the action by terms linear in the fields.

The Ward identities related to extended Galilean invariance allow to express 1PI correlation functions of order $(n+1)$ with one velocity field having vanishing wave vector in terms of 1PI correlation functions of order $n$. This is an exceptional phenomenon for statistical field theories and has been studied previously in the context of fluid turbulence [56, 61–63]. A particularly useful consequence in the context of the functional renormalisation group is that flow equations can be closed in certain limits. To this end, it has been argued that velocity fluctuations are in fact the leading contribution to the flow of two-point functions in the limit of large wave numbers, being the limit in which flow equations can indeed be formally closed. One should be cautious at this point since there might be non-perturbative features, such as additional scales set by e.g. non-vanishing velocity dispersion, which can invalidate the argument and modify the large wave number behaviour of correlation functions. It would be highly interesting to investigate from an observational point of view whether this could be used as a possible signal for non-trivial features in the statistical properties of dark matter.

From a physical point of view it was argued that the suppression of unequal-time correlation functions at large wave numbers is related to what has been called sweeping effect in the context of fluid turbulence [65]. Density fluctuations on small scales are transported by a large-scale velocity field in a random way, leading to a decorrelation with time. However, in contrast to a truly dissipative suppression, correlation functions at equal times are not affected.

The present paper focussed on conceptual developments but it also lays the ground for further investigations. Specifically, the functional renormalisation group equations can be solved with numerical methods when truncating the space of 1PI effective actions. This allows for non-perturbative computations of correlation functions, such as the propagator and power spectrum, and one expects useful insights in particular for time and length scales where perturbative methods fail. It is regarded as particularly interesting to develop a complete understanding of non-linear cosmological structure formation, also at relatively small scales where velocity dispersion and other effects like dark matter self-interactions could start to play a role. The latter can be included by working with the Boltzmann equation rather than with its collisionless limit and approximating the collision term in an appropriate manner, e.g. the Stoßzahlansatz.

## Appendix A: Galilean invariance and non-renormalisation of the gravitational sector

Instead of eliminating the gravitational potential by solving Poisson's equation (4), one can equivalently use a gravitational response field $\hat{\phi}(\tau, \boldsymbol{x})$ in order to enforce Poisson's equation. Constraint equations are introduced into the generating functional (14) in the same manner as field equations so that the bare action in this setting reads

$$\begin{aligned} S[\psi,\hat{\psi},\phi,\hat{\phi}] = &- \mathrm{i} \int_{\tau,\boldsymbol{x},\boldsymbol{x}'} \hat{\psi}_a(\tau,\boldsymbol{x}) \Big[\partial_\tau \delta_{ab}\, \delta(\boldsymbol{x}-\boldsymbol{x}') + \varOmega'_{ab}(\tau,\boldsymbol{x}-\boldsymbol{x}')\Big] \psi_b(\tau,\boldsymbol{x}') \\ &- \mathrm{i} \int_{\tau,\boldsymbol{x},\boldsymbol{x}',\boldsymbol{x}''} \hat{\psi}_a(\tau,\boldsymbol{x})\, \gamma_{abc}(\boldsymbol{x}-\boldsymbol{x}',\boldsymbol{x}-\boldsymbol{x}'')\, \psi_b(\tau,\boldsymbol{x}')\, \psi_c(\tau,\boldsymbol{x}'') \\ &+ \int_{\boldsymbol{x},\boldsymbol{x}'} \hat{\psi}_a(\tau_{\mathrm{in}},\boldsymbol{x}) \Big[\mathrm{i}\, \delta(\boldsymbol{x}-\boldsymbol{x}')\, \varPsi_a^{\mathrm{in}} + \tfrac{1}{2}\, C_{ab}^{\mathrm{in}}(\boldsymbol{x}-\boldsymbol{x}')\, \hat{\psi}_b(\tau_{\mathrm{in}},\boldsymbol{x}')\Big] \\ &- \mathrm{i} \int_{\tau,\boldsymbol{x}} \hat{u}_i(\tau,\boldsymbol{x})\, \partial_i \phi(\tau,\boldsymbol{x}) - \mathrm{i} \int_{\tau,\boldsymbol{x}} \hat{\phi}(\tau,\boldsymbol{x}) \Big[\partial_i\partial_i \phi(\tau,\boldsymbol{x}) - \tfrac{3}{2}\, \mathcal{H}^2 \varOmega_{\mathrm{m}} \delta(\tau,\boldsymbol{x})\Big] . \end{aligned} \tag{A1}$$

Here, $\varOmega'_{ab}$ is the upper triangular part of the matrix $\varOmega_{ab}$ given in equation (9), that is the velocity-density com-ponent due to integrating out the gravitational potential is removed. This has the advantage that no non-analyticities are present in the bare action since $\Omega'_{ab}$ only acts through spatial gradients.

The gravitational sector of the theory is particularly simple since there are two extended symmetries related to the (infinitesimal) time- and space-gauged field shifts $\delta_\epsilon \phi(\tau, \boldsymbol{x}) = \epsilon(\tau, \boldsymbol{x})$ and $\delta_\epsilon \hat{\phi}(\tau, \boldsymbol{x}) = \epsilon(\tau, \boldsymbol{x})$. These yield the Ward identities

$$\Gamma^{(1,0)}_{k,\phi}(\tau, \boldsymbol{x}) = \mathrm{i}\left[\partial_i \hat{u}_i(\tau, \boldsymbol{x}) - \partial_i \partial_i \hat{\phi}(\tau, \boldsymbol{x}')\right] , \tag{A2}$$

and

$$\Gamma^{(0,1)}_{k,\phi}(\tau, \boldsymbol{x}) = -\mathrm{i}\left[\partial_i \partial_i \phi(\tau, \boldsymbol{x}) - \tfrac{3}{2}\,\mathcal{H}^2 \Omega_\mathrm{m} \delta(\tau, \boldsymbol{x})\right] , \tag{A3}$$

which encode that the whole gravitational sector is not renormalised and the dependence on the gravitational fields is the same for the bare and effective action.

In this setting Galilean invariance can be realised as a true symmetry using the transformations (45) in addition to

$$\begin{aligned} \delta_\epsilon \phi(\tau, \boldsymbol{x}) &= -\epsilon_i(\tau)\,\partial_i \phi_a(\tau, \boldsymbol{x}) \\ &\qquad - x_i[\ddot{\epsilon}_i(\tau) + \mathcal{H}\dot{\epsilon}_i(\tau)] , \\ \delta_\epsilon \hat{\phi}(\tau, \boldsymbol{x}) &= -\epsilon_i(\tau)\,\partial_i \hat{\phi}_a(\tau, \boldsymbol{x}) . \end{aligned} \tag{A4}$$

In the case where the gravitational potential is integrated out using Poisson's equation (4),

$$\partial_i \phi(\tau, \boldsymbol{x}) = \int_{\boldsymbol{x}'} O_i(\tau, \boldsymbol{x} - \boldsymbol{x}')\,\delta(\tau, \boldsymbol{x}') + C_i(\tau, \boldsymbol{x}) , \tag{A5}$$

where the operator $O_i$ is defined in equation (10), one has a residual gauge symmetry due to the arbitrary solenoidal vector field $C_i$. By choosing $C_i$ appropriately, any bulk velocity terms appearing due to a Galilean transformation (45) can be eliminated. Since $C_i = 0$ is fixed in the equations of motion (8), Galilean invariance is no longer manifest. This should be understood as 'gauge fixing' to the frame in which the velocity mean field is vanishing.[20]

## Appendix B: 1PI two-point function flow equations in the large external wave number limit

In the following, the first diagram of the flow equation (34) is explicitly computed in the large external wave number limit. The other diagrams of the flow equations (34) and (35) are evaluated in a similar fashion to arrive at the flow equations (63) and (64). The first flow diagram of equation (34) is given by

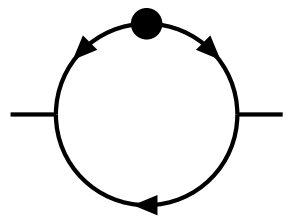

$$\begin{aligned} &= -\mathrm{i}\,\bar{\delta}(\boldsymbol{q} + \boldsymbol{q}') \int_{\substack{\bar{\tau}, \bar{\tau}' \\ \tilde{\tau}, \tilde{\tau}'}} \int_{\boldsymbol{l}} \hat{\Gamma}^{(2,1)}_{k,cea}(\tilde{\tau}, \boldsymbol{l}; \bar{\tau}, -\boldsymbol{q} - \boldsymbol{l}; \tau, \boldsymbol{q})\, G^\mathrm{R}_{k,ef}(\bar{\tau}, \bar{\tau}', \boldsymbol{q} + \boldsymbol{l})\, \hat{\Gamma}^{(2,1)}_{k,dbf}(\tilde{\tau}', -\boldsymbol{l}; \tau', -\boldsymbol{q}; \bar{\tau}', \boldsymbol{q} + \boldsymbol{l}) \\ &\qquad \times \hat{\partial}_k P^\mathrm{I}_{k,dc}(\tilde{\tau}', \tilde{\tau}, \boldsymbol{l}) \\ &\sim -\mathrm{i}\,\bar{\delta}(\boldsymbol{q} + \boldsymbol{q}') \int_{\substack{\bar{\tau}, \bar{\tau}' \\ \tilde{\tau}, \tilde{\tau}'}} \hat{\Gamma}^{(2,1)}_{k,u_j ea}(\tilde{\tau}, \boldsymbol{0}; \bar{\tau}, -\boldsymbol{q}; \tau, \boldsymbol{q})\, G^\mathrm{R}_{k,ef}(\bar{\tau}, \bar{\tau}', \boldsymbol{q})\, \hat{\Gamma}^{(2,1)}_{k,u_i bf}(\tilde{\tau}', \boldsymbol{0}; \tau', -\boldsymbol{q}; \bar{\tau}', \boldsymbol{q}) \\ &\qquad \times \hat{\partial}_k \int_{\boldsymbol{l}} P^\mathrm{I}_{k,u_i u_j}(\tilde{\tau}', \tilde{\tau}, \boldsymbol{l}) \\ &= -\mathrm{i}\,\bar{\delta}(\boldsymbol{q} + \boldsymbol{q}') \int_{\bar{\tau}, \bar{\tau}'} D^\mathrm{R}_{k,ae}(\tau, \bar{\tau}, \boldsymbol{q})\, G^\mathrm{R}_{k,ef}(\bar{\tau}, \bar{\tau}', \boldsymbol{q})\, D^\mathrm{R}_{k,fb}(\bar{\tau}', \tau', \boldsymbol{q}) \\ &\qquad \times \frac{q^2}{3} \int_{\tilde{\tau}, \tilde{\tau}'} \left[\theta(\tilde{\tau} - \bar{\tau}) - \theta(\tilde{\tau} - \tau)\right]\left[\theta(\tilde{\tau}' - \tau') - \theta(\tilde{\tau}' - \bar{\tau}')\right] \hat{\partial}_k \int_{\boldsymbol{l}} P^\mathrm{I}_{k,u_i u_i}(\tilde{\tau}', \tilde{\tau}, \boldsymbol{l}) \\ &= -\mathrm{i}\,\bar{\delta}(\boldsymbol{q} + \boldsymbol{q}')\, q^2 \int_{\bar{\tau}, \bar{\tau}'} D^\mathrm{R}_{k,ae}(\tau, \bar{\tau}, \boldsymbol{q})\, G^\mathrm{R}_{k,ef}(\bar{\tau}, \bar{\tau}', \boldsymbol{q})\, D^\mathrm{R}_{k,fb}(\bar{\tau}', \tau', \boldsymbol{q})\, \hat{\partial}_k I_k(\tau, \bar{\tau}; \bar{\tau}', \tau') , \end{aligned} \tag{B1}$$

[20] Within the functional integral representation (14) a specific velocity mean field can be forced by evaluating expectation values at a non-vanishing response field source current or by adding a 'frame-fixing' term to the bare action [58]. Similar to local gauge symmetries the choice of frame can be gauge fixed using the Faddeev–Popov method so that one obtains a Becchi–Rouet–Stora symmetry and a related Slavnov–Taylor identity instead of an extended Galilean symmetry [69, 70].

where the second equality holds in the limit $q \to \infty$ under the assumptions discussed in section V. The third equality makes use of the Ward identity (61) and statistical isotropy implies

$$q_i q_j \int_{\boldsymbol{l}} P^{\mathrm{I}}_{k,u_i u_j}(\tau, \tau', \boldsymbol{l}) = \frac{q^2}{3} \int_{\boldsymbol{l}} P^{\mathrm{I}}_{k,u_i u_i}(\tau, \tau', \boldsymbol{l}) \ . \tag{B2}$$

Finally, the last equality uses definition (65) in order to rewrite the expression.

## ACKNOWLEDGMENTS

This work is supported by the Deutsche Forschungsgemeinschaft (German Research Foundation) under Germany's Excellence Strategy and the Cluster of Excellence EXC 2181 (STRUCTURES), the Collaborative Research Centre SFB 1225 (ISOQUANT) as well as the research grant FL 736/3-1.